\documentclass[12pt]{iopart}

\usepackage{iopams}
\usepackage[USenglish]{babel} 
\usepackage{dsfont}
\usepackage{rotating}
\usepackage{xcolor}
\usepackage{amsmath}
\usepackage{amssymb}
\usepackage{amsthm}
\usepackage[urlcolor=blue,colorlinks=true,linkcolor=blue,pdfstartview={FitH},bookmarks=false]{hyperref} 
\usepackage{mciteplus} 
\newcommand{\ket}[1]{\ensuremath{|#1\rangle}}
\newcommand{\bra}[1]{\ensuremath{\langle#1|}}

\newcommand{\braket}[2]{\ensuremath{\langle#1|#2\rangle}}

\newcommand{\eg}{\emph{e.g.}}
\newcommand{\ie}{\emph{i.e.}}
\newcommand{\etal}{\emph{et al}}

\newcommand{\p}{\scriptscriptstyle{+}}
\newcommand{\m}{\scriptscriptstyle{-}}

\newcommand{\dg}{\dagger}

\newcommand{\mc}{\mathcal}
\newcommand{\mb}{\mathbb}
\newcommand{\ot}{\otimes}
\newcommand{\op}{\oplus}

\newcommand{\tx}[1]{\text{#1}}
\newcommand{\txt}[1]{\text{#1}}

\mciteSetMidEndSepPunct{\newline\hspace*{-0.3cm}}{.}{\relax}

\begin{document}

\title{Initial states of qubit--environment models leading to conserved quantities}

\author{Bart{\l}omiej Gardas}
\address{Institute of Physics, University of Silesia, PL-40-007 Katowice, Poland}
\ead{bartek.gardas@gmail.com}

\author{Jerzy Dajka}
\address{Institute of Physics, University of Silesia, PL-40-007 Katowice, Poland}
\ead{jerzy.dajka@us.edu.pl}

\begin{abstract}
It is possible to prepare a composite qubit--environment system so that its time evolution will
guarantee the conservation of a preselected qubit's observable. In general, this observable is 
not associated with a symmetry. The latter may not even be present in the subsystem. The initial
states which lead to such a quantity conserved dynamics form a subspace of the qubit--environment
space of states. General construction of this subspace is presented and illustrated by two examples.
The first one is the exactly soluble Jaynes--Cummings model and the second is the multi--photon 
Rabi model.   
\end{abstract}

\maketitle

\section{Introduction}
An interaction between two quantum systems almost always modifies a set of quantum numbers suitable for
the description of non--interacting components. Generically, only in the absence of interactions the set
of good quantum numbers of a composite system consists of good quantum numbers of its subsystems. Let
us consider the energy as an example. It is well--known that (Hamiltonian--type) interactions cause an
energy flow from one system to the other and/or vice--verse. The total system remains conservative but 
the interacting parts become {\it open} which changes their character qualitatively~\cite{breu}. 

One can name at least one case, commonly referred to as {\it dephasing}~\cite{cd,*dd,*pd,*nd,*od,*adc}, 
where the energy exchange between two systems is absent during the entire evolution regardless of the 
initial conditions. In open quantum system theory, where it is often assumed that the second subsystem 
is much larger than the first one, this phenomenon is also known as the pure decoherence~\cite{SB_Alicki,*pud,*dajka,*dg}. 
It induces the loss of quantum coherence (and, as a result, the emergence of classical behaviour~\cite{dmwter}) 
without affecting system's energy.

For dephasing to occur a rather specific type of the interaction is required, yet nothing which might
contradict our intuition is behind this phenomenon. The energy lossless evolution is due to the existence
of a conservation law, the total Hamiltonian commutes with the interaction operator. It is also the case
if, instead of the energy, one considers a quantum number related to a local observable (\ie{} acting 
nontrivially only on a subsystem). In general, one cannot expect such a number to be conserved unless 
the corresponding observable happens to be a symmetry in that subsystem. 

One can ask: Is it possible to design such a time evolution of a composite quantum system that makes a 
local quantum number a good quantum number regardless of the existence of the related symmetry in the 
subsystem? This is very intriguing as we are really asking whether the conservation of a quantum number 
in an open system is possible not despite destructive influence of its environment but rather because of 
it. At face value, this may sound paradoxically (to say the least), yet similar behaviour has been shown
to take place in \eg{} quantum Markovian systems~\cite{cirac,*dt}.

Such a behaviour might serve as a model of a different type of computation, dissipative quantum computation, 
in which it is the environment which does all the work. Instead of isolating a system (and then applying a 
tractable unitary dynamics on it) as in the standard paradigm of quantum computation~\cite{fd}, we engineering
its coupling with the environment to do the desire tasks~\cite{qs,*ee,t1,*t2,*qmd} (also thermodynamics 
once~\cite{cbh}). In a scenario like this, the interaction is describe by a Markovian master equation~\cite{alicki}, 
the computation task is encoded in Lindblad operators, which model the environment, whereas the unique outcome
(results of the computation) is written into the steady state reached rapidly by the open system. 

This work is devoted to show that the destructive nature of the environment can also be explored to induce 
conservation of information encoded in open quantum systems. We only consider two--level systems, the general
case remains open. There are many potential application of this mechanism. Take for example quantum information 
science. 

Quantum devices of the future, like \eg{} quantum computers~\cite{QCNature}, will most likely be composed of components
consist of a large amount of qubits (quantum memory~\cite{QMNature}, quantum register~\cite{Register2,*QRNature}, etc.). 
The challenge is not only to stabilize such devices (shield against decoherence) but also to program them. For instance, 
from various reasons (\eg{} to establish a reference point or to cache the results) it may be desired to froze the spin 
of a single qubit (or a cluster of them) in a given (preselected) direction during computational cycles. 
In principle, this could be achieved by means of quantum control usually via suitable drivings. Instead of (either open
loop or feedback) quantum control, which continuously affects the system being analyzed, we propose a method in which it
is sufficient to measure the system only once at beginning of the evolution (initial preparation). No dynamical control is 
needed to maintain the desired dynamics. Such alternative approach can be more beneficial as the goal can be achieved without 
introducing `numerical' errors resulting from the measurement. Of course, to fully explore this idea in real environments
much work both experimental and theoretical needs to be done. We hope that this paper will serve as a simple theoretical 
starting point.

The layout of this paper is as follows. We begin by showing how a proper choice of the initial state of a composite 
qubit--boson system can assure no energy exchange between the two subsystems. In other words, we identify dephasing--like 
behaviour in non--dephasing model. Next, in section~\ref{two}, we generalize this idea further, beyond the energy, by showing
how to prepare a qubit--environment system so that the information encoded in a preselected qubit observable cannot be erased
by its environment. Section~\ref{three} followed by the conclusions~\ref{four} serves as our second example.

\section{An example: Conservation of the energy}
\label{one}
Let us start with the Jaynes--Cummings (JC) model~\cite{vedral},
\begin{equation}
\label{jm}
\begin{split}
\tx{H}&=\omega\sigma_z+\nu a^{\dagger}a+\left(g^*\sigma_{+}\otimes a+g\sigma_{-}\otimes a^{\dagger}\right).      
\end{split}
\end{equation}
$a$ and $a^{\dagger}$ are the creation and annihilation operators of the bosonic field (environment) \ie, 
$[a,a^{\dagger}]=\mathbb{I}_{\tx{B}}$, whereas $\sigma_z$ denotes the one of the Pauli spin matrices with
eigenstates $\ket{\pm}$. $\sigma_{\pm}$ are ladder operators, $\sigma_{\pm}\ket{\mp}=\ket{\pm}$ and 
$\sigma_{\pm}\ket{\pm}=0$. $\nu$ and $\omega$ describe energies of the qubit and the field, respectively.
The coupling constant $g$ reflects the strength of the interaction between the systems.

Although the model~(\ref{jm}) essentially originates from quantum optics~\cite{vedral}, it has been studied in a wide
range of other branches of physics (see \eg~\cite{jco,*jch,*jcm}) for almost half--century within the broad variety of
contexts~\cite{JCS,*jc_breuer,*jc_breuer2,*jc_kwek}. Here, it serves as a good starting point for our considerations. 

In this example, we are interested in finding a qubit--boson initial state, $\varrho$, such that the qubit energy, 
$E_{\tx{Q}}=\omega\tx{Tr}\left(\sigma_{z}\otimes\mathbb{I}_{\tx{B}}\varrho(t)\right)\sim \langle \sigma_z\rangle$,
is conserved. In other words, we wish to have $E_{\tx{Q}}=\omega\tx{Tr}\left(\sigma_{z}\rho(t)\right)=cst.$ where 
$\rho(t)=\tx{Tr}_{\tx{B}}(e^{-i\tx{H}t}\varrho e^{i\tx{H}t})$ stands for the reduced qubit dynamics~\cite{alicki}.

It is well known that for the JC model the total number of excitation, $\tx{N}=a^{\dagger}a+\sigma_z$, is conserved. 
This fact ultimately leads to the exact diagonalization of the JC Hamiltonian (the exact dynamics of the JC model is
known). At this point, we want to emphasize that none of the above features is essential in the following analysis, 
yet involving $\tx{N}$ one simplifies calculations.

We start by splitting the Hamiltonian~(\ref{jm}) into two commuting parts, $\tx{H}=\tx{H}_0+\tx{V}$, where
$\tx{H}_0 = \nu\tx{N}$ and
\begin{equation}
\label{split}
\tx{V} = \delta\sigma_z+\left(g^*\sigma_{+}\otimes a+g\sigma_{-}\otimes a^{\dagger}\right).
\end{equation}
$\delta\equiv\omega-\nu$ is the detuning frequency. As a result, the time evolution of the system can be factorized 
such that
\begin{equation}
\label{esplit}
\tx{U}_t=\exp(-i\tx{H}_0t)\exp(-i\tx{V}t)\equiv\tx{U}_t^0\tx{V}_t.
\end{equation}
Note, $\rho(t)$ mimics a pure dephasing evolution if 
\begin{equation}
\label{defaz}
\rho(t)=
\alpha |+\rangle\langle+|+ c(t)^*|-\rangle\langle+| 
+c(t)|+\rangle\langle-| + (1-\alpha)|-\rangle\langle-|,
\end{equation}
where $\alpha$ is a real constant and $c(t)$ denotes a function of time such that $|c(t)|\le 1$.
So, the questions are: Which initial states of the composite system do guarantee the reduced 
dynamics~(\ref{defaz})? Are they separable or entangled?

In order to give an answer to this questions, we offer an explicit construction of such states. At face value, subsequent
steps may seem bizarre and to some extent artificial. Notwithstanding this, they can be performed, as it is discussed below,
for a broad class of qubit--boson models. As a first step, we define $\varrho=\ket{\Psi}\bra{\Psi}$ where
\begin{equation}
\label{init}
\ket{\Psi}=C_{\psi}\left( \ket{+}\otimes\ket{\psi}+\ket{-}\otimes\tx{X}\ket{\psi} \right),
\end{equation}
and $\ket{\psi}$ is an arbitrary state of the bosonic field. $C_{\psi}$ is a normalization constant chosen so that
$\langle\Psi|\Psi\rangle=1$. For the sake of simplicity, we absorb it into $\ket{\psi}$ so that $\|\psi\|=|C_{\psi}|$.

A possibility of constructing dephasing states relies on the ability of finding a linear operator $\tx{X}$ satisfying 

\begin{equation}
\label{ricc}
 g^*\tx{X}a\tx{X}+2\delta\tx{X}-ga^{\dagger}=0,
\end{equation}
or $\tx{X}(\delta+g^*a\tx{X})=-\delta\tx{X}+g a^{\dagger}$. As it will be seen, this choice of $\tx{X}$ allows for a 
specific time evolution of the above state, namely 
\begin{equation}
\label{later}
\ket{\Psi}\rightarrow\ket{\Psi_t}=
\tx{U}_t^0\left(\ket{+}\otimes\ket{\psi_t}+\ket{-}\otimes\tx{X}\ket{\psi_t}\right).
\end{equation} 
According to Eq.~(\ref{later}), the dynamics of the entire system consists of two parts. The one governed by
$\tx{U}_t^0=\exp(-i\tx{H}_0t)$ is the free evolution. The other one, encoded in $\ket{\psi_t}$, is the
remaining part resulting from the interaction. It needs to be determined along with $\tx{X}$. As a result 
of the evolution~(\ref{later}), we obtain the following reduced dynamics

\begin{equation}
\label{ac}
\alpha(t)=\braket{\psi_t}{\psi_t}
\quad\txt{and}\quad
c(t)=e^{-i2\nu t}\bra{\psi_t}e^{i\nu a^{\dagger}at}\tx{X}e^{-i\nu a^{\dagger}at}\ket{\psi_t},
\end{equation}
which resembles the dephasing dynamics~(\ref{defaz}) if the vector $\ket{\psi_t}$ itself evolves unitarily. 
For this to hold true there must exist a Kamiltonian $\tx{K}$ such that \mbox{$\ket{\psi_t}=e^{-i\tx{K}t}$\ket{\psi}}.
To see this is indeed the case here let's assume that $\tx{K}=\delta+g^*a\tx{X}$. Then it follows from~(\ref{ricc}) 
and~(\ref{split}) that

\begin{equation}
\label{v}
\begin{split}
\tx{V}\ket{\Psi}
&=
\ket{+}\otimes\tx{K}\ket{\psi}
+
\ket{-}\otimes\left(-\delta\tx{X}+g a^{\dagger}\right)\ket{\psi} \\
&=
\ket{+}\otimes\tx{K}\ket{\psi}
+
\ket{-}\otimes\tx{XK}\ket{\psi},
\end{split}
\end{equation}
which holds for every $\ket{\psi}$. Thus, by replacing $\ket{\psi}$ with $\tx{K}\ket{\psi}$ one finds
\begin{equation}
\tx{V}^2\ket{\Psi}=\ket{+}\otimes\tx{K}^2\ket{\psi}+\ket{-}\otimes\tx{X}\tx{K}^2\ket{\psi}.
\end{equation}
Repeating this step $n$ times we have
\begin{equation}
\label{vn}
\tx{V}^n\ket{\Psi}
=
\ket{+}\otimes\tx{K}^n\ket{\psi}
+
\ket{-}\otimes\tx{X}\tx{K}^n\ket{\psi}
\end{equation}
and finally
\begin{equation}
\label{vt}
\tx{V}_t\ket{\Psi}
=
\sum_{n=0}^{\infty}\frac{(-it)^n}{n!}\tx{V}^n\ket{\Psi} 
=
\ket{+}\otimes e^{-i\tx{K}t}\ket{\psi}
+
\ket{-}\otimes \tx{X}e^{-i\tx{K}t}\ket{\psi}.
\end{equation}
It remains `only' to identify $\tx{X}$ and to show that $\tx{K}$ is Hermitian. There is no general method 
allowing to find solutions of the quadratic equations~(\ref{ricc}), yet judging from its structure it seems
reasonable to try $\tx{X}=\sum_n\xi_n\ket{n+1}\bra{n}$. One can infer that this is indeed a good guess as long as
\begin{equation}
\xi_n=\frac{-\delta+\sqrt{\delta^2+|g|^2(n+1)}}{g^*\sqrt{n+1}},
\quad n\geq 0.
\end{equation}
The solution we have just found generalizes the Susskind--Glogower operator~\cite{glow} in the sense that
$\tx{X}\rightarrow(aa^{\dagger})^{-1/2}a^{\dagger}$ when $\delta\rightarrow 0$. As a result, 
$\tx{K}=\sqrt{\delta^2+|g|^2(a^{\dagger}a+\mathbb{I}_{\tx{B}})}$ and it clearly is Hermitian. 
It follows from Eq.~(\ref{ac}) that $\alpha=\braket{\psi}{\psi}$ and in addition 
\begin{equation}
c(t)=e^{-i\nu t}\sum_{n=0}^{\infty}\xi_ne^{i\Omega_n t}\braket{\psi}{n+1}\braket{n}{\psi},
\end{equation}
where $\Omega_n=\sqrt{\delta^2+|g|^2(n+2)}-\sqrt{\delta^2+|g|^2(n+1)}$.

Note, we still have the freedom to choose the vector $\ket{\psi}$. In particular, putting $\ket{\psi}\sim\ket{m}$,
where $\ket{m}$ is a state with defined number of bosons, we not only have $\dot{c}(t)=0$ (meaning $\rho(t)$ is a 
steady state) but also $c(t)=0$ (\ie{} the state exhibits its classical nature). 

In general, states like the one from Eq.~(\ref{init}) are entangled. Hence, there is no one--to--one correspondence 
between qubit--boson density operators \ket{\Psi}\bra{\Psi} and reduced density matrices
$\tx{Tr}_{\tx{B}}(\ket{\Psi}\bra{\Psi})$ of the qubit~\cite{korelacje,*erratum}. If, however, $\ket{\psi}$ is an 
eigenvector of $\tx{X}$, that is $\tx{X}\ket{\psi}=\lambda\ket{\psi}$ for $\lambda\in\mathbb{C}$, then
$\ket{\Psi}=\left(\ket{+}+\lambda\ket{-}\right)\otimes{\ket{\psi}}$ is separable.


\section{Generalization}
\label{two}
So far, we have showed how to determine initial preparations of a (specific) composite qubit--boson system which 
guarantee no energy flow between the qubit and its environment (boson). In that case, the reduced qubit dynamics 
mimics a pure dephasing evolution. We have also investigated conditions under which from this broad class of states 
one can choose separable ones. 

Now, we are interested in finding an answer to a more general question: How to prepare an initial state of a composite
system (and determine its separability) which result in no `information flow' between its subsystems. We assume that 
the information is encoded in a qubit observable $\Lambda$. As we will see, such states have very much in common with 
dephasing states and for that reason we will keep this terminology.

In section~\ref{one}, we have conducted our analysis by making use of a very specific exactly solvable model. Currently, 
we will show that neither solvability nor the form of the interaction in this model is a {\it sine qua non} condition for 
building `dephasing states' for general qubit--environment models.  

Let $\Lambda$ be a $2\times 2$ Hermitian matrix--a given qubit observable. Our objective is to determine 
$\rho\equiv\rho(0)$ such that $\langle\Lambda(t)\rangle\equiv\tx{Tr}\left(\Lambda\rho(t)\right)$ remains
constant during the evolution. As before, by $\rho(t)$ we denote the qubit reduced dynamics,
$\rho(t)=\tx{Tr}_{\tx{E}}\left[e^{-i{\bf{H}}t}\ket{\Psi}\bra{\Psi} e^{i{\bf{H}}t}\right]$, where
$\ket{\Psi}$ is the initial qubit--environment state and 
\begin{equation}
\label{husula}
\tx{\bf{H}}=\tx{H}_{\tx{Q}}\otimes\mathbb{I}_{\tx{E}}+\mathbb{I}_{\tx{Q}}\otimes\tx{H}_{\tx{E}}+\tx{\bf{H}}_{\tx{int}}.
\end{equation}
with all the symbols having their usual meaning, stands for the total Hamiltonian.

We begin with a very simple observation that in each moment of time and for every complex $2\times 2$ matrix
$\Lambda$ the partial trace, $\tx{Tr}_{\tx{E}}(\cdot)$, satisfies 

\begin{equation}
\tx{Tr}\left[\Lambda \tx{Tr}_{\tx{E}}(\varrho(t))\right]=\tx{Tr}\left[(\Lambda\otimes\mathbb{I}_{\tx{E}})\ket{\Psi(t)}\bra{\Psi(t)}\right].
\end{equation}
By virtue of this relation, we have
\begin{equation}
\label{av}
 \langle\Lambda(t)\rangle=
 \tx{Tr}\left[\left(\Lambda_d\otimes\mathbb{I}_{\tx{E}}\right)\ket{\Omega(t)}\bra{\Omega(t)}\right],
\end{equation}
%
where $\ket{\Omega(t)}=e^{-i\tx{\bf{K}}t}\ket{\Omega}$ with $\ket{\Omega}=\tx{U}\otimes\mathbb{I}_{\tx{E}}\ket{\Psi}$ and
\begin{equation}
\label{trans}
{\bf{K}}=(\tx{U}\otimes\mathbb{I}_{\tx{E}}){\bf{H}}(\tx{U}\otimes\mathbb{I}_{\tx{E}})^{\dagger}.
\end{equation}
$\tx{U}$ denotes the unitary matrix such that $\tx{U}^{\dagger}\Lambda\tx{U}\equiv\Lambda_{\tx{d}}=\tx{diag}(\lambda_{\p},\lambda_{\m})$.

The Kamiltonian $\bf{K}$ can always be written as 
\begin{equation}
\label{hgen}
\tx{\bf{K}} =\ket{+}\bra{+}\otimes\tx{H}_{\p}+\ket{-}\bra{-}\otimes\tx{H}_{-} 
            +\ket{+}\bra{-}\otimes\tx{V}+\ket{-}\bra{+}\otimes\tx{V}^{\dagger}.       
\end{equation}
An explicit form of $\tx{H}_{\pm}$ and $\tx{V}$ can easily be recovered when the Hamiltonians $\tx{H}_{\tx{Q}}$, $\tx{H}_{\tx{E}}$, 
and $\bf{H}_{\tx{int}}$ are provided. {\it A priori}, we neither impose any physical restriction of their specification nor 
assume the existence of symmetries in the total system. 

The composite system is assumed to be in the state $\varrho=\ket{\Psi}\bra{\Psi}$ initially, where
\begin{equation}
\label{ginit}
\ket{\Psi}=C_{\psi}\left(\ket{\lambda_{\p}}\otimes\ket{\psi}+\ket{\lambda_{\m}}\otimes\tx{X}\ket{\psi}\right),
\end{equation}
with $\ket{\lambda_{\pm}}=\tx{U}^{\dagger}\ket{\pm}$. $\ket{\psi}$ is a freely chosen state of the environment.
As before, one can redefine the state $\ket{\psi}$ so that $\ket{\psi}\rightarrow C_{\psi}\ket{\psi}$. 
We show that $\ket{\Omega}$ undergoes the following evolution

\begin{equation}
\label{glater}
\ket{\Omega}
\rightarrow
e^{-i{\bf{K}}t}\ket{\Omega}=\ket{+}\otimes\ket{\psi_t}+\ket{-}\otimes\tx{X}\ket{\psi_t},
\end{equation} 
where $\ket{\psi_t}=e^{-i\tx{K}_{\p}t}\ket{\psi}$ for some operator $\tx{K}_{\p}$, provided that $\tx{X}$ solves 
the (operator) Riccati equation,
\begin{equation}
\label{gricc}
\tx{XVX}+\tx{XH}_{\p}-\tx{H}_{\m}\tx{X}-\tx{V}^{\dagger}=0.
\end{equation}
$\tx{K}_{\p}$ and thus $\ket{\psi_t}$ are yet to be determined. Riccati equations of the type~(\ref{gricc}) not only 
have been studied in mathematical contexts~\cite{ricc_book} but also have recently been applied to investigate open 
quantum systems~\cite{Rdiag,*gardas3,*gardas4}. As one may anticipate, Eq.~(\ref{gricc}) reduces to the condition~(\ref{ricc}) 
when $\bf{H}$ is given by~(\ref{jm}).  

It immediately follows from~(\ref{gricc}) that $\tx{X}(\tx{H}_{\p}+\tx{VX})=\tx{H}_{\m}\tx{X}+\tx{V}^{\dagger}$ 
which leads to
\begin{equation}
{\bf{K}}\ket{\Omega}=\ket{+}\otimes\tx{K}_{\p}\ket{\psi}+\ket{-}\otimes\tx{X}\tx{K}_{\p}\ket{\psi}
\quad\txt{where}\quad
\tx{K}_{\p}=\tx{H}_{\p}+\tx{VX}.
\end{equation}
As easily as before, one can justify the general formula,
\begin{equation}
\label{hn}
{\bf{K}}^n\ket{\Omega}=\ket{+}\otimes\tx{K}_{\p}^n\ket{\psi}+\ket{-}\otimes\tx{X}\tx{K}_{\p}^n\ket{\psi}
\quad (n\geq 0).
\end{equation}
As a result, the evolution generated by $\bf{K}$ reads 
\begin{equation}
\label{gvt}
\ket{\Omega(t)}
=
\sum_{n=0}^{\infty}\frac{(-it)^n}{n!}{\bf{K}}^n\ket{\Psi} 
=
\ket{+}\otimes e^{-i\tx{K}_{\p}t}\ket{\psi}
+
\ket{-}\otimes \tx{X}e^{-i\tx{K}_{\p}t}\ket{\psi}.
\end{equation}

 

Having~(\ref{gvt}) in place, we can begin to investigate conditions upon which the expectation value~(\ref{av}) does not depend
on time. Except that $\alpha$ is time--dependent, $\txt{Tr}_{\txt{E}}(\ket{\Omega(t)}\bra{\Omega(t)})$ has a structure which is
similar to the dephasing matrix~(\ref{defaz}). To be more specific, 
\begin{equation}
\label{dd}
\alpha(t)=\bra{\psi}e^{i\txt{K}_{\p}^{\dg}t}e^{-i\txt{K}_{\p}t}\ket{\psi} 
\quad\txt{and}\quad
c(t)=\bra{\psi}e^{i\txt{K}_{\p}^{\dg}t}\txt{X}e^{-i\txt{K}_{\p}t}\ket{\psi}.
\end{equation}
These formulas generalize those given in Eq.~(\ref{ac}) as one should expect.

Similar dynamics can also be designed by starting from any state orthogonal to~(\ref{ginit}).
Indeed, such orthogonal states are found to be of the form
\begin{equation}
\label{ort}
\ket{\Phi}=C_{\phi}\left(\ket{\lambda{\m}}\otimes\ket{\phi}-\ket{\lambda{\p}}\otimes\tx{X}^{\dagger}\ket{\phi}\right),
\end{equation}
which can be verified by noticing that $\langle\Psi|\Phi\rangle=0$ for every $\ket{\psi}$, $\ket{\phi}\in\mathcal{H}_{\tx{E}}$. 
In this case, the qubit's reduced dynamics takes the form
\begin{equation}
\label{odd}
\alpha(t)
=1-\braket{\phi_t}{\phi_t},\quad c(t)=-\bra{\phi_t}\txt{X}^{\dagger}\ket{\phi_t}, 
\quad
\ket{\phi_t}=e^{-i\txt{K}_{\m}t}\ket{\phi}.
\end{equation}
where $\tx{K}_{\m}=\tx{H}_{\m}-\tx{V}^{\dagger}\tx{X}^{\dagger}$. For instance, in the Jaynes--Cummings model discussed in the
preceding section we have $\tx{K}_{\m}=\sqrt{\delta^2+|g|^2a^{\dagger}a}$.

Henceforward, we assume that the composite system is prepared initially in one of the states~(\ref{defaz}).
Then it follows from Eq.~(\ref{av}) 

\begin{equation}
 \langle\Lambda(t)\rangle=\alpha(t)\lambda_{\p}+\left(1-\alpha(t)\right)\lambda_{\m}.
\end{equation}
Clearly, if $\alpha(t)$ is a constant so is $\langle\Lambda(t)\rangle$. Thus the question is: Upon which conditions can we
have $\alpha(t)=cst.$? Obviously, it depends on particular properties of $\txt{K}_{\p}$. So, what are they? First of all, if 
$\txt{VX}$ is Hermitian, as in our opening example, so is $\txt{K}_{\p}$ and hence the evolution $\ket{\psi}\rightarrow\ket{\psi_t}$
is unitary. Thus $\alpha(t)=\alpha(0)$ and yet $c(t)\not=c(0)$ which result in not trivial dephasing dynamics regardless of 
the initial vector $\ket{\psi}$.

The really interesting question here is: What about all the cases when $\txt{K}_{\p}$ is not Hermitian? Are those even
possibly? It is hard to address these questions since we have no knowledge regarding $\txt{X}$. Nevertheless, with a fair
amount of intuition combined with justifiable assumptions concerning the operator $\txt{K}_{\p}$ we can overcome this problem. 

First, we argue that $\txt{K}_{\pm}$ are pseudo--Hermitian \ie, there are some invertible $\eta$ and $\xi$ such that 
$\txt{K}_{\p}^{\dg}=\eta\txt{K}_{\p}\eta^{-1}$ and $\txt{K}_{\m}^{\dg}=\xi\txt{K}_{\m}\xi^{-1}$~\cite{ali12}. Indeed, if 
$\txt{X}$ solves the Riccati Eq.~(\ref{gricc}) then (note $\mb{C}^2\ot\mc{H}_{\txt{E}}=\mc{H}_{\txt{E}}\op\mc{H}_{\txt{E}}$)

\begin{equation}
\label{sim}
\begin{bmatrix}
1 & -\txt{X}^* \\
\txt{X} & 1
\end{bmatrix}^{-1}
\begin{bmatrix}
\txt{H}_{\p} & \txt{V} \\
\txt{V}^{\dagger} & \txt{H}_{\m}
\end{bmatrix}
\begin{bmatrix}
1 & -\txt{X}^* \\
\txt{X} & 1
\end{bmatrix}
=
\begin{bmatrix}
\txt{K}_{\p} & 0 \\
0 & \txt{K}_{\m}
\end{bmatrix},
\end{equation}
meaning, in particular, that $\sigma({\bf{K}})=\sigma(\txt{K}_{\p})\cup\sigma(\txt{K}_{\m})$. Moreover, if one defines 
$\bf{U}$ to be the unitary matrix from the polar decomposition of $\bf{S}$, $\bf{S}=\bf{U}\sqrt{\bf{S}^{\dg}\bf{S}}$, 
then it follows form~(\ref{sim}) that

\begin{equation}
\label{izo}
{\bf{U}}^{\dg}
\begin{bmatrix}
\txt{H}_{\p} & \txt{V} \\
\txt{V}^{\dagger} & \txt{H}_{\m}
\end{bmatrix}
{\bf{U}}
=
%
({\bf{S}}^{\dg}{\bf{S}})^{1/2}
\begin{bmatrix}
\txt{K}_{\p} & 0 \\
0 & \txt{K}_{\m}
\end{bmatrix}
({\bf{S}}^{\dg}{\bf{S}})^{-1/2}.
\end{equation}
Since a unitary transformation preserves Hermiticity, the latter equality proves the pseudo--Hermicity conditions 
given above for $\eta=\mb{I}_{\txt{E}}+\txt{X}^{\dg}\txt{X}$ and $\xi=\mb{I}_{\txt{E}}+\txt{X}\txt{X}^{\dg}$, respectively.
It should be obvious that $\eta$ and $\xi$ induce positive--defined inner products $\langle\eta\cdot,\cdot\rangle$ and
$\langle\xi\cdot, \cdot\rangle$ (because of that they are called metric operators) with respect to which $\txt{K}_{\p}$ and
$\txt{K}_{\m}$ are Hermitian, respectively.

It seams reasonable to assume that the Kamiltonian $\txt{K}_{\p}$ has a complete set of eigenstates. For the sake of argument,
let's also assume that its spectrum is discrete and not degenerated (\ie{} $\txt{K}_{\p}\ket{\psi_n}=E_n\ket{\psi_n}$). Then
\begin{equation}
\txt{K}_{\p}=\sum_{n}E_n\ket{\psi_n}\bra{\psi_n}
\quad\txt{and}\quad
\txt{K}_{\p}^{\dg}=\sum_{n}E_n^*\ket{\psi_n}\bra{\psi_n}.
\end{equation}
Now, the interesting part follows: According to~(\ref{sim}), an eigenvalue $E_n$ of $\txt{K}_{\p}$ is also an eigenvalue of 
$\bf{K}$. Thus $E_n^*=E_n$ proving that $\txt{K}_{\p}$ is in fact Hermitian. 

If one cannot find a basis consists with eigenstates of the Kamiltonian $\txt{K}_{\p}$ we still can design a dephasing 
dynamics as long as $\txt{K}_{\p}$ is diagonalizable. This condition is much weaker than Hermicity and it seems to reflect
an absolute minimal physical requirement one can impose on $\txt{K}_{\p}$ in this context~\cite{ali0}.

As before, we only investigate discrete and not degenerated case. Saying that $\txt{K}_{\p}$ can be diagonalized means that
there is a basis $\{\ket{n}\}$, linear invertible transformation $\txt{S}$ and complex numbers $E_n$ (which we known are in 
fact real) such that
\begin{equation}
 \txt{S}^{-1}\txt{K}_{\p}\txt{S}=\sum_{n}E_n\ket{n}\bra{n}.
\end{equation}
Now, we can introduce two sets of vectors: $\ket{\psi_n}=\txt{S}\ket{n}$ and $\ket{\phi_n}=(\txt{S}^{-1})^{\dg}\ket{n}$ 
which form a complete set of biorthonormal eigenvectors~\cite{ali13,*aliBat}, that is

\begin{equation}
\label{bi}
\txt{K}_{\p}\ket{\psi_n}=E_n\ket{\psi_n},
\quad
\txt{K}_{\p}^{\dg}\ket{\phi_n}=E_n\ket{\phi_n}, 
\quad
\mb{I}_{\txt{E}}=\sum_{n}\ket{\psi_n}\bra{\phi_n},
\quad
\langle\psi_n|\phi_m\rangle =\delta_{nm}.
\end{equation}
%
At this point, it should be stressed that $\langle\psi_n|\psi_m\rangle\not=\delta_{nm}$ and
$\langle\phi_n|\phi_m\rangle\not=\delta_{nm}$ in general. In view of~(\ref{bi}), we have

\begin{equation}
\txt{K}_{\p}=\sum_{n}E_n\ket{\psi_n}\bra{\phi_n}
\quad\txt{and}\quad
\txt{K}_{\p}^{\dg}=\sum_{n}E_n\ket{\phi_n}\bra{\psi_n}.
\end{equation}
Note, if the similarity operator $\txt{S}$ is unitary so is the evolution $\ket{\psi}\rightarrow\ket{\psi_t}$ since
the two basis $\{\ket{\psi_n}\}$,$\{\ket{\psi_m}\}$ are identical and thus $\txt{K}_{\p}$ is Hermitian. Henceforward,
we assume this is not the case because we have already examined it. Form both~(\ref{dd}) and~(\ref{bi}) we have

\begin{equation}
\alpha(t)
=
\sum_{n}|\braket{\psi}{\phi_n}|^2\|\psi_n\|^2 
+
\sum_{n\not=m}e^{i(E_n-E_m)t}\braket{\psi}{\phi_n}\braket{\psi}{\phi_m}^*\langle\psi_n|\psi_m\rangle 
\end{equation}
and in addition to that

\begin{equation}
c(t)
=
\sum_{n}|\braket{\psi}{\phi_n}|^2\|\txt{X}\psi_n\|^2 
+
\sum_{n\not=m}e^{i(E_n-E_m)t}\braket{\psi}{\phi_n}\braket{\psi}{\phi_m}^*\langle\psi_n|\txt{X}|\psi_m\rangle. 
\end{equation}

Both this expression are time--dependent as they should (in a Hilbert space one cannot have a linear and non 
unitary evolution which preserves the norm). We can, however, makes $\alpha(t)$ constant at least for certain
initial states $\ket{\psi}$. In particular, if $\ket{\psi}=\ket{\psi_n}$ then both $\alpha(t)$ and $c(t)$ does
not depend on time. The latter observation is in perfect agreement with the results regarding stationary states 
reported in~\cite{gardasPuchala}.


\section{Second example}
\label{three}
As a second example, we consider the $k$-photon Rabi model~\cite{rabi_Nstates,*srwa} for which $\tx{H}_{\tx{Q}}=\omega\sigma_z$
and $\tx{H}_{\tx{E}}=\nu a^{\dagger}a$ read exactly as in the James--Cummings model~(\ref{jm}) but the interaction is given by
\begin{equation}
 \label{rabbi}
 {\bf{H}}_{\tx{int}}=\sigma_x\otimes(g^*a^{k}+g(a^{\dg})^k).
\end{equation}
This model not only generalizes the single mode case but also includes counter--rotating--wave terms~\cite{crt,*crt2,*ctr3}, which
make its analytical treatment much more complicated in compare with~(\ref{jm}). Despite recent progress in finding it analytic 
solution~\cite{braak,*braak_pra,*zieg}, the problem remains highly non--trivial.

Let us suppose that this time we want to find initial state(s) of the composite system such that the $x$--component
of the qubit spin operator remains constant during the evolution, $S_{x}=\tfrac{1}{2}\tx{Tr}\left(\sigma_{x}\rho(t)\right)=cst.$

First, one needs to determine $\tx{U}$ which transforms $\sigma_x$ to its diagonal form. It is an easy task
to do and the answer is $\tx{U}=\left(\sigma_z+\sigma_x\right)/\sqrt{2}$. In this case, we have 
$\tx{U}\sigma_x\tx{U}^{\dagger}=\sigma_z$, that is $\lambda_{\pm}=\pm 1$. Next, we transform $\bf{H}$ into $\bf{K}$ 
according to~(\ref{trans}) and then we recover $\tx{H}_{\pm}$, $\tx{V}$ from the decomposition~(\ref{hgen}). 
Straightforward calculation shows

\begin{equation}
\label{krabih}
\tx{H}_{\pm}=\nu a^{\dg}a\pm \left(g^*a^{k}+g(a^{\dg})^k\right),
\quad
\tx{V}=\omega \mb{I}_{\tx{E}}. 
\end{equation}
The corresponding Riccati equation~(\ref{gricc}) reads as follows
\begin{equation}
\label{gricc1}
\omega\tx{X}^2+\tx{X}\tx{H}_{\p}-\tx{H}_{\m}\tx{X}-\omega\mathbb{I}_{\tx{E}}=0.
\end{equation}
We can solve this equation by introducing the generalize parity operator~\cite{multi}: 
\begin{equation}
\label{sol}
 \tx{X}_k=\sum_{l=1}^{k}\sum_{n=0}^{\infty}(-1)^n\ket{n,l}\bra{n,l}, 
\end{equation}
where $\ket{n,l}:=\ket{kn+l-1}$ and $l\leq k$. This operator is both Hermitian and unitary 
(in particular $\tx{X}_k^2=\mathbb{I}_{\tx{E}}$). For $k=1$ it simplifies to the well--known
bosonic parity $\tx{P}=\exp(i\pi a^{\dagger}a)$. As one can verify, $\tx{X}_ka^{k}\tx{X}_k=-a^{k}$
and thus $\tx{X}_k\tx{H}_{\p}\tx{X}_k=\tx{H}_{\m}$. Therefore, $\tx{X}_k$ indeed solves~(\ref{gricc1}). 
This fact may come as a surprise since~(\ref{sol}) does not depend on any of the parameters $\nu$, $\omega$, $g$. 

In this model, the dephasing dynamics~(\ref{dd}) and~(\ref{odd}) are generated by 
$\tx{K}_{\pm}=\tx{H}_{\pm}\pm\omega\tx{X}_k$, respectively. Introducing projections
$\tx{P}_{\pm}=\tfrac{1}{2}(\mathbb{I}_{\tx{E}} \pm \tx{X}_k)$ onto subspaces
$\mathcal{H}_{\pm}$ consist of states with defined parity (with respect to the generalized 
parity~(\ref{sol})) and taking into account both~(\ref{init}) and~(\ref{ort}), we have 
\begin{equation}
\ket{\Psi_{\epsilon}}=\frac{1}{2}\left(\ket{+}\otimes\tx{P}_{\epsilon}\ket{\psi}+\ket{-}\otimes\tx{P}_{-\epsilon}\ket{\psi}\right),
\quad\epsilon=\pm1,
\end{equation}
which are separable if $\ket{\psi}\in\mathcal{H}_{\pm}$. 

\section{Summary}
%
\label{four}
The more conservation laws are present in a system (either classical or quantum) the better is our understanding of
its behaviour and properties. In this paper, we have presented a method of designing a time evolution of a composite 
qubit--environment system which guarantees the conservation of a preselected qubit's observable (its energy for instance) 
in the absence of the related conservation law. This can be done by a proper choice of the initial state of the total
system (dephasing state). We have also argued that such initial states are entangled in general, yet one can disentangle
them in principle. 
  
Besides obvious applications, it may be also possible to apply our results either for testing the precision of initial
state preparation of qubit–-boson systems or for examining if a chosen Hamiltonian properly describes a quantum system.
The conservation of $\langle\Lambda\rangle$ is granted if one successfully prepares the initial state $\ket{\Psi}$ from 
Eq.~(\ref{defaz}) (or~Eq.~(\ref{ort})). On the other hand, the dynamics of $\langle\Lambda\rangle$ can serve as a `first
test', usually simpler than \eg{} tomographic methods, for the quality of applied state engineering. Having prepared 
the right initial state, any departure from $\langle\Lambda\rangle=cst.$ can indicate either significant influence of 
noise (decoherence) or necessity of modifying the system Hamiltonian (by including non--linear terms, for instance).

To select the right initial state a solution of the Riccati equation is required. Although neither the form nor even 
the existence of such a solution can be taken for granted in general, some useful criteria of solvability, applicable 
to a broad range of relevant physical systems, can be found in literature~\cite{Vadim,*Vadim2,*langer}. Nowadays when 
powerfully computers and accurate numerical methods are accessible, it is a secondary issue, to say the least, that we
cannot solve this equation analytically.
 
A Preparation of the observable--conserving states $\ket{\Psi}$ (or $\ket{\Phi}$) would require highly sophisticated quantum
engineering. One could try to construct the state $\ket{\chi}=\tx{X}\ket{\psi}$, then `tensorize' it with $\ket{-}\otimes\ket{\chi}$
and finally superpose the result with the separable qubit--environment state $\ket{+}\otimes\ket{\psi}$. Clearly, the situation 
becomes simpler if the state which we try to prepare is separable as it is when $\ket{\psi}$ is an eigenstate of $\tx{X}$.

Our description of `state preparation' is extremely naive and it does not even pretend to be an experimental suggestion, especially 
when the initial qubit--environment state is entangled. This is very serious limitation, but our work is purely theoretical and any
deeper analysis of experimental perspectives is essentially beyond its scope. Despite this difficulties, we hope that together with
continuous development of quantum state engineering techniques the construction proposed in this work can become useful for applications.


\ack
This work was supported by the Polish National Science Center under grants
UMO-2011/01/N/ST3/02473 (B. G) and N202 052940 (J. D)

\section*{References}


\providecommand{\newblock}{}

\end{document}